\documentclass[prl,aps,twocolumn, floatfix, superscriptaddress,showpacs]{revtex4}
\usepackage{amsmath,bm,graphicx}
\usepackage{amssymb}
\usepackage{amsfonts}
\begin{document}

\title{Decorrelating a compressible turbulent flow: an experiment}

\author{Jason Larkin}
\email[Corresponding Author : ]{jml37+@pitt.edu }
\affiliation{Department of Physics \& Astronomy, University of Pittsburgh, Pittsburgh, PA 15260, USA.}

\author{Walter I. Goldburg}
\affiliation{Department of Physics \& Astronomy, University of Pittsburgh, Pittsburgh, PA 15260, USA.}

\date{\today}

\begin{abstract}
Floating particles that are initially distributed uniformly on the surface
of a turbulent fluid, subsequently coagulate, until  finally a steady state is reached.  This being so, they manifestly form a compressible system.  In this experiment, the information dimension $D_1$, and the Lyapunov exponents  of the coagulated floaters, is measured.  The trajectories and the velocity fields of the particles are captured in a sequence of rapidly acquired images. Then the velocity sequence is randomly shuffled in time to generate new trajectories.  This analysis mimics the Kraichnan ensemble and yields properties of a velocity correlation function that is delta-correlated in time (but not in space).  The measurements are compared with theoretical expectations and with  simulations of Boffetta et al., that closely mimic the laboratory experiment reported here.
\end{abstract}


\maketitle
\section{Introduction}
 It has become a standard procedure to study fluid flows by tracking the trajectories of small particles introduced into the flow.  Assume first that  the fluid is highly viscous and moves azimuthally between two concentric cylinders, one of which is rotating at constant angular velocity. A series of consecutive and rapid velocity measurements will reveal that each particle is moving in a circle.   Now imagine that the series of consecutive velocity measurements is randomly rearranged, as one might shuffle a deck of cards.   If the shuffled velocity field is then used to evolve tracer particles (explained in the Experiment section), they will appear to be jerking back and forth.  Nonetheless,  the  circularity of the trajectories will be recognizable.   Now imagine that the fluid is turbulent rather than laminar and that the  frame rate of the camera (in Hz) is much faster than the inverse lifetime of the smallest eddies.  Will the randomly shuffled sequence of measurements contain information about the turbulence, as in the laminar case?

 This laboratory study was motivated by a computer simulation of this same problem by Boffetta et al. (to be called BDES) \cite{BDES2004}.
 That simulation, in turn, was inspired by a theoretical analysis of velocity fluctuations by R. Kraichnan \cite{kraichnan1968}.  Concentrating on the two-point velocity autocorrelation function,  Kraichnan argued that one could increase our understanding of  turbulence by  making the simplifying assumption that velocity fluctuations are delta-correlated in time, while retaining the spatial dependence of the fluctuations.  Kraichnan's approach is related to prior work of L. F. Richardson, who analyzed the time evolution of the separation of particle pairs in turbulent flows  by treating their motion as Brownian-like \cite{RICH1926}.

This experiment, and  the computer simulations of BDES,  focuses on an unusual type of flow, namely the motion of low-density particles that  \underline {float} on a turbulent, three-dimensional incompressible fluid (water in these experiments). Even though the underlying fluid is incompressible, the floating particles form a compressible system. If the water molecules should move downward at some point at the surface ($z$ = 0), the floaters, which cannot follow this downward motion, will accumulate there.  Likewise they will flee the upwelling points.  Since the underlying flow is assumed to be turbulent, the pattern the floaters form will fluctuate  in both time and space.  The floaters are chosen to be small enough that they follow the velocity field at the surface; their inertia is negligible (see below).  Neglecting the wave motion of the surface (which is negligible), the floaters move in a plane.  They are merely sampling the velocity field of the incompressible flow at the surface, and hence their velocity at each instant $t$ is the same as that of the surface water molecules $(v_x(x,y,0,t), v_y(x,y,0,t))={\bf v}({\bf x},t))$. The geometrical pattern formed by the floaters has a fractal appearance, as seen in Fig. \ref{particle} (a), which reaches a statistical steady state in approximately 1s.  This work focuses on the geometric and dynamic behavior of the floaters in the steady state.

The coagulation phenomenon studied here and in BDES has nothing to do with the interaction between the floaters or the waves that are present on the surface; all the effects observed in the laboratory are also seen in computer simulations, where these two effects are not included \cite{CDGS2004,BDES2004}.   If the flow field is isotropic, one can define a dimensionless compressibility at the surface,
\begin{equation}
\label{compressibility}
{\cal C}  =  \frac{\langle  (\partial_x v_x + \partial_y v_y )^2  \rangle}{2\langle  ( \partial_x  v_x - \partial_y  v_y)^2      \rangle}
\end{equation}
which  lies between 0 (incompressible flow) and 1( potential flow).  In the simulations ${\cal C}$ is an adjustable parameter, but in the experiments it is measured to be very close to 1/2 \cite{CDGS2004}.  The computer simulations yield a value very close to this \cite{BDES2004,CDGS2004}. In these experiments a descriptor of the  coagulation pattern is extracted from the data,  namely the ``information dimension'' $D_1$, which is defined below. It is simply related to ${\cal C}$ in the Kraichnan model \cite{BDES2004}.  Thus, the measurement of $D_1$ provides an experimental test of the applicability of the Kraichnan model for temporally decorrelated data.

\section{Experiment}
\begin{figure}[h]
\begin{center}$
\begin{array}{cc}
\includegraphics[width=2.5 in]{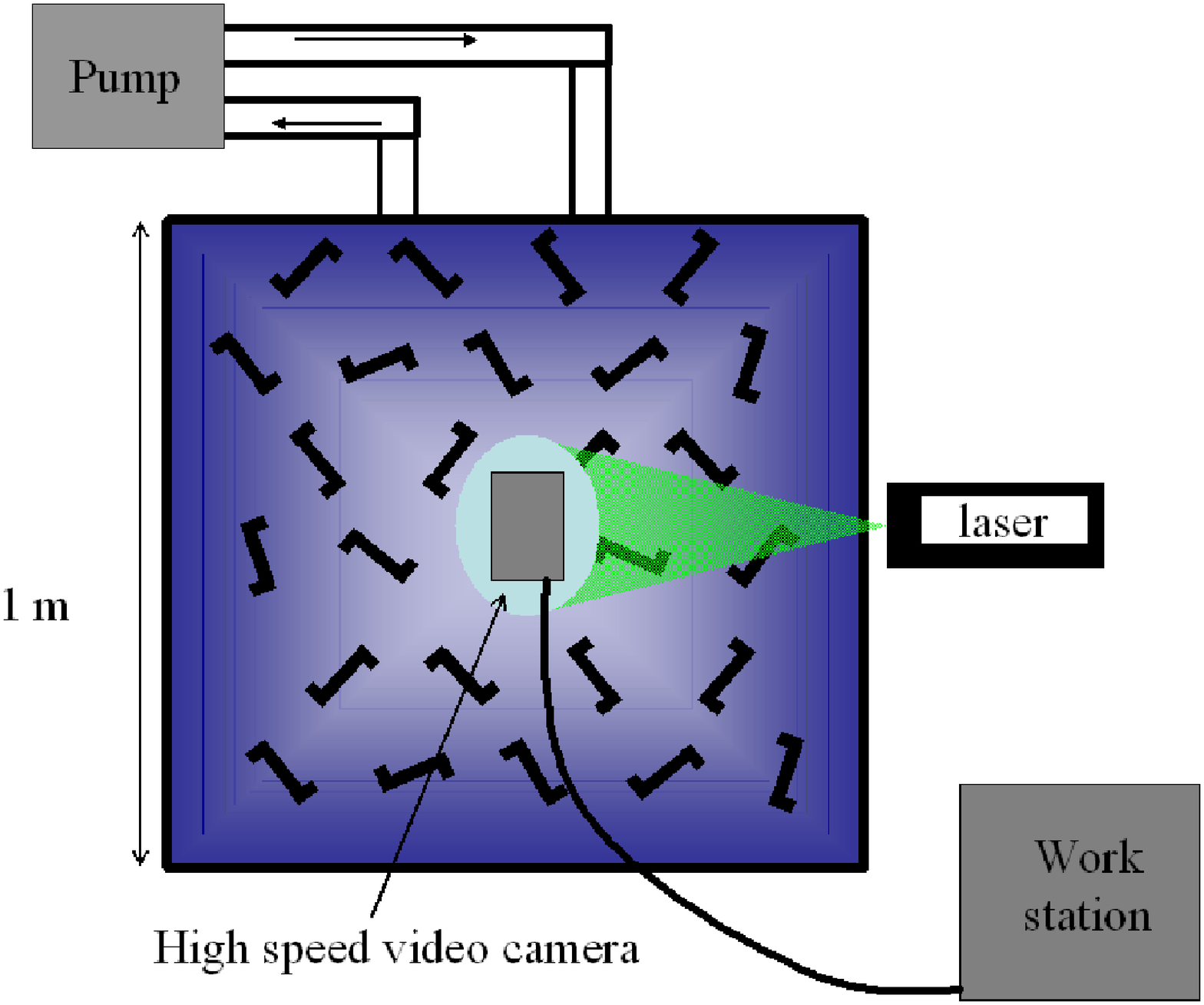} &\\
\includegraphics[width=2.5 in]{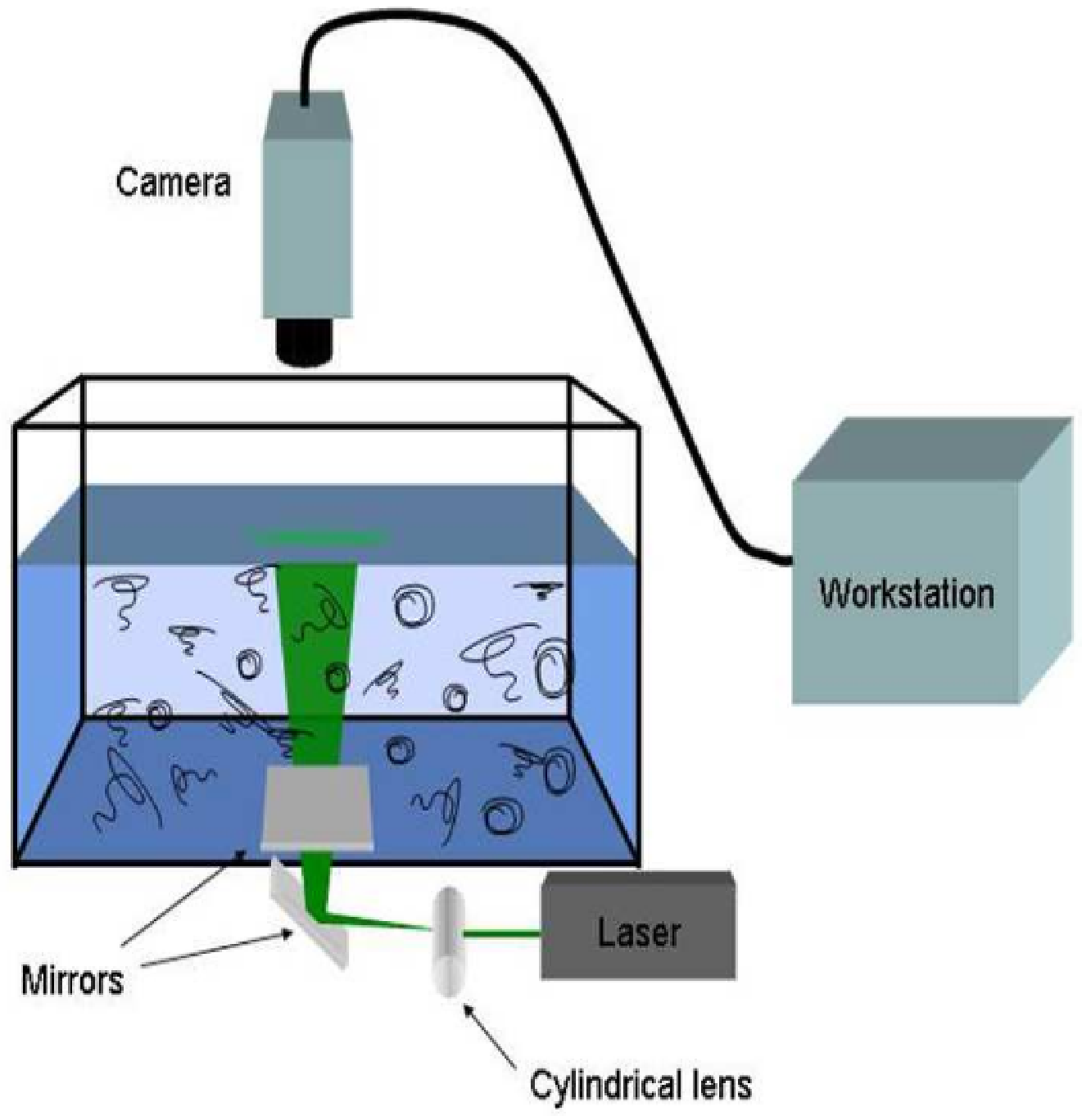}
\end{array}$
\end{center}
\caption{Schematic of the top-view (top panel) and side-view (bottom panel) of the experimental setup. 36 rotating capped jets are placed horizontally on the tank floor (shown as randomly oriented Z-shaped patterns) that pump water into the tank re-circulated by a 8hp pump. The central region of the water surface is illuminated by a laser-sheet. A high-speed digital camera suspended vertically above this central region captures images of the light scattered by buoyant particles (50 $\mu$m hollow-glass spheres of specific gravity 0.25).}
\label{setup}
\end{figure}

The experiments are carried out in a tank 1m $\times$ 1m in lateral
dimensions, filled with water to a depth of $30 cm$. Turbulence is
generated by a pump ($8 hp$) which re-circulates water through a system of 36 rotating jets placed horizontally across the
tank floor (see Fig. \ref{setup}). This system for generating the turbulence ensures that the source of turbulent injection is far
removed from the free surface where the measurements are made \cite{CDGS2004}. More
importantly, the method also minimizes the amplitude of surface waves, which are unavoidable.  Prior experiments have demonstrated that the waves at the surface have negligible influence on the experimental results \cite{CDGS2004}.

The hydrophilic particles chosen here are subject to capillary forces which
are very small compared to forces coming from the turbulence, and do not affect the results as they do in \cite{PFL2006, FWDL2005}. The non-inertial character of the particles is minimal because the product $\tau_s \lambda_1$, where $\lambda_1$ is the largest Lyapunov exponent and $\tau_s$ is the stopping time of the particle \cite{BCG2008}, is of the order $5 \times 10^{-4}$, which is too small to lead to any significant inertial effect \cite{CDGS2004}.

During an experimental run, floating particles ($ 50 \mu m$ diameter and specific gravity of 0.25) are constantly seeded into the fluid from the tank floor, where they undergo turbulent mixing as they rise due to buoyancy. Their motion is tracked via a high-speed camera (Phantom v.5) situated above the tank. The camera field-of-view is a square area of side length $L = 9 cm$. The constant particle injection is necessary to replace particles at the surface during the experiment. The source and sink structure at the surface fluctuates in both time and space, which can cause particles to leave the camera's field of view. This experimental scheme has been used several times before \cite{CDGS2004,bandi2006}, including a detailed comparison of the experiment with numerical simulations.

The instantaneous velocity field of the floaters is measured using an in-house developed particle imaging velocimetry (PIV) program, which processes the recorded images of the surface particles. The constant injection of particles ensures that surface sources and sinks receive an adequate coverage on the surface. The local particle density at the surface determines the average spacing of the velocity vector fields produced by the PIV program. The resulting velocity vectors are spaced (on average) by $\delta x =$ 2.5 $\eta$ over both sources and sinks, where $\eta$ is the dissipative scale at the surface (Table I).   This density of the vector grid spacing is important for the Lagrangian particle evolution scheme, which is discussed below.

The camera height above the water surface was chosen so that a pixel size is
roughly $0.1 mm$, comparable to $\eta$ (see Table I).
Data were taken for several values of $Re_\lambda \simeq 150-170$ (defined in Table I) with an average $Re_\lambda \simeq 160$. The measurements in this experiment show no systematic variation with the $Re_\lambda$ over this range, so they are used as an ensemble average so as to decrease measurement errors. Turbulent parameters measured at the surface are listed in Table \ref{table}. The camera frame rate is set between 100-200 $Hz$, such that each instantaneous velocity vector field is separated by approximately $4-8 \tau_{\eta}$, where $\tau_{\eta}$ is the lifetime of eddies of size $\eta$ (see Table \ref{table}) . Results reported here do not depend on this frame rate.

The experimentally measured velocity fields were then used to
solve the equation of motion for Lagrangian particles :
\begin{equation}
\label{advection}
\frac{d{\bf x}_i}{dt} = {\bf v}({\bf x}_i(t),t),
\end{equation}
where ${\bf v}({\bf x}_i,t)$ is the velocity field and ${\bf x_i}=(x_i,y_i)$ are the individual particle positions.

To achieve accurate results for the Lagrangian particle evolution, the vector fields used in Eq. \ref{advection} were interpolated from the experimentally determined velocity vectors via a
bi-cubic interpolation scheme developed for numerical simulations, as discussed in \cite{Pope1988} and implemented in \cite{CDGS2004}. This scheme uses the smooth flow between grid points separated by length scales comparable to $\eta$ to interpolate the velocity field between measured grid points.  To use this scheme it is necessary for the measured grid spacing to satisfy the criterion $\delta x < \pi \eta$, where $\delta x$ is the measured velocity grid spacing. We have tested to ensure that the results do not depend on the velocity grid spacing by varying the spacing from $\delta x$ = 2.5 $\eta$ to $\delta x$ = 4 $\eta$ (increasing this average spacing even further degrades the statistical quality of the data).

\begin{figure}[h]
\begin{center}$
\begin{array}{cc}
\includegraphics [width=2.5 in, height=2.63 in]{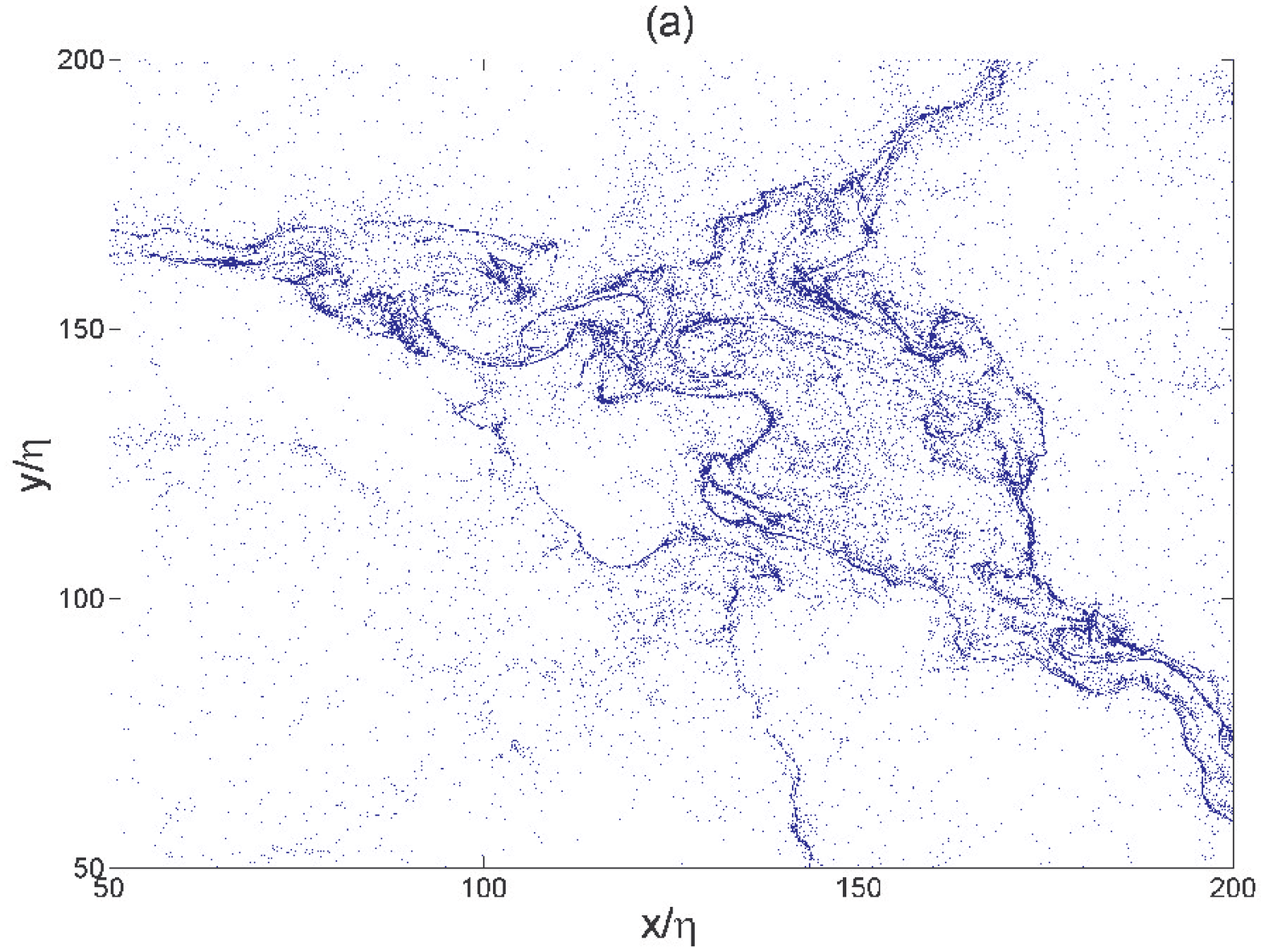} &\\
\includegraphics [width=2.5 in, height=2.63 in]{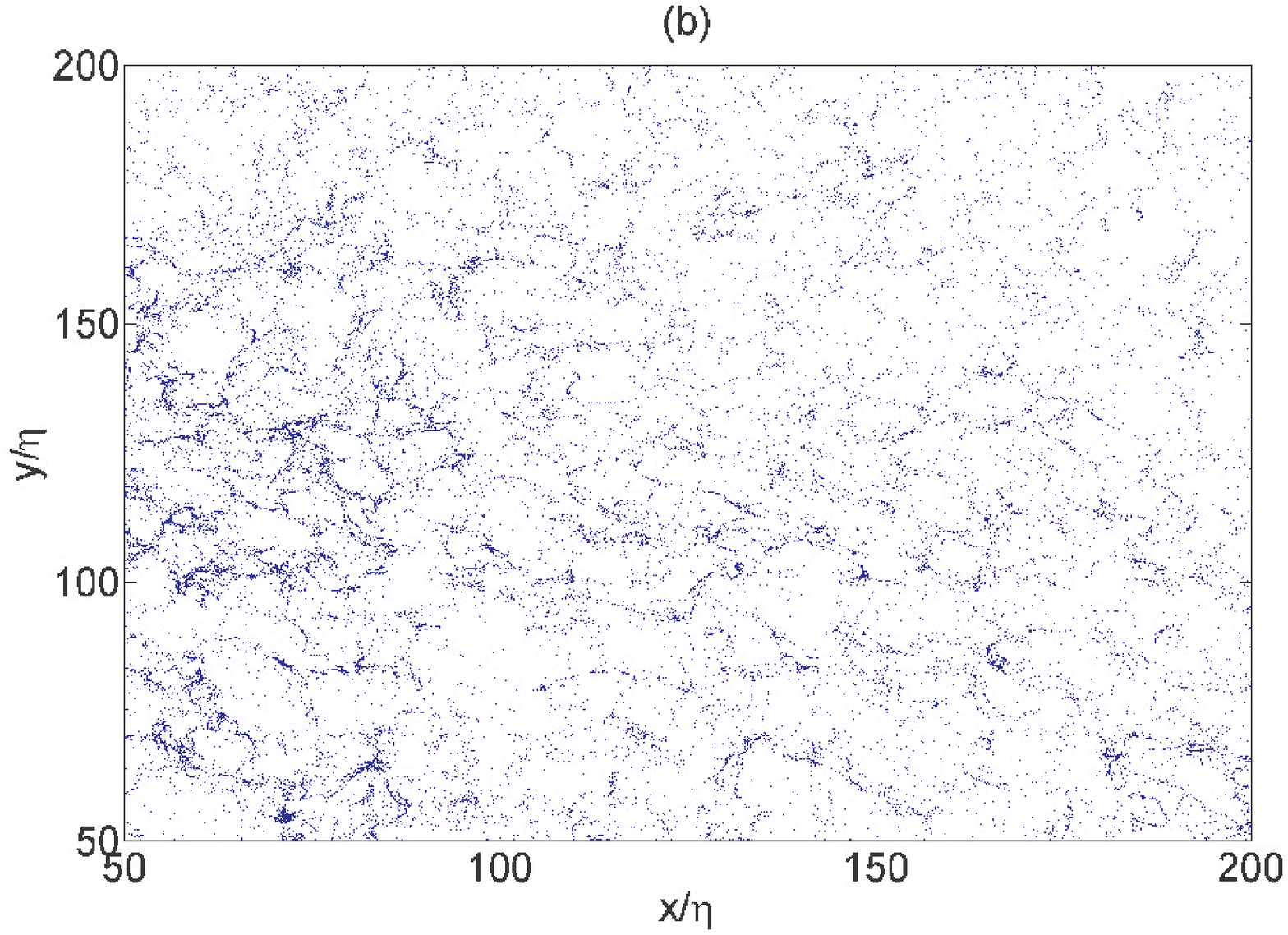}
\end{array}$
\end{center}
\caption{Visualization of particle distributions for the correlated flow (a) and the decorrelated flow (b) at $t/\tau_0=5$. The time scale is in units of the large eddy turnover time $\tau_0$. The initial distribution ($t/\tau_0=0$) is homogeneous with 4$\times 10^5$ particles.  Both particle distributions are in the steady state. Notice that for the correlated flow, there are coherent structures at large scales ($r > 50$). For the decorrelated flow, these large scale structures are not present, but line-like coagulations persist at scale $r < 10$.}
\label{particle}
\end{figure}

For the correlated flow data, a uniform distribution of Lagrangian tracers is generated at time $ t/\tau_0 = 0$, where $\tau_0 = \delta U (l_0) / l_0 $ is the large eddy turnover time. Here, $\tau_0 = 0.4 s$. The uniform distribution is then evolved using the above-described scheme with the velocity vector field record in the order it was taken during the experiment. To de-correlate the data, we take successive snapshots of the Eulerian velocity field and put them in random order. An initially uniform particle distribution is then evolved from this shuffled record. The velocity vector field record spans roughly 60 $\tau_{0}$. The results in this work were unaffected when this record was cut in half, suggesting that the data record is long enough to mimic delta correlation in time. Visualizations of the correlated and decorrelated particle distributions (in the steady state) are shown in Fig. \ref{particle}(a) and (b) respectively. The qualitative impression from the two images is that both distributions lie along line-like structures, but for the correlated flow these structures span up to several integral length scales, $l_0/\eta \simeq 70$, where $l_0$ is defined in Table I.

\begin{table}
\caption{\label{table}Turbulent parameters measured at the surface.  Measurements are made at several values of the $Re_\lambda$ with an average $Re_\lambda \simeq 160$.  The parameters listed are averages, with deviations less than $10\%$.}
\begin{center}
\begin{tabular}{|p{1.2in}|p{1.2in}|p{0.5in}|}
 \hline
 \small {Taylor microscale} & $\lambda $=$\sqrt{\frac{v^{2} _{rms} }{\left\langle \left({\partial v_{x}
\mathord{\left/{\vphantom{\partial v_{x}  \partial x}}\right.\kern-\nulldelimiterspace} \partial x} \right)^{2}
\right\rangle } }$ & 0.47 (cm) \\ \hline
\small{Taylor Re$_\lambda$} & Re$_\lambda$=$\frac{v_{rms} \lambda }{\nu }$ & 160 \\ \hline
\small{Integral Scale} & $l_0$=$\int dr\frac{\left\langle v_{\left\| \right. }
(x+r)v_{\left\| \right. } (x)\right\rangle }{\left\langle
\left(v_{\left\| \right. } (x)\right)^{2} \right\rangle }$ & 1.42 (cm) \\ \hline
    \small{Large Eddy Turnover Time} & $\tau _{0} =\frac{l_{0} }{v_{rms} } $ & 0.43 (s) \\ \hline
    \small{Kolmogorov Time} & $\tau_{\eta}$=$\frac{\nu}{\epsilon}^{1/2}$ & 0.04 (s) \\ \hline
    \small{Dissipation Rate} & $\varepsilon_{diss}$=$10\nu \left\langle \left(\frac{\partial v_{x} }{\partial x} \right)^{2} \right\rangle $ & 6.05 ($cm^2/s^3$) \\ \hline
    \small{Kolmogorov scale} & $\eta =\left(\frac{\nu ^{3} }{\varepsilon }\right)^{1/4} $ & 0.02 (cm) \\ \hline
    \small{RMS Velocity} & $v_{rms} =\sqrt{\left\langle v^{2} \right\rangle-\left\langle v\right\rangle ^{2} }$ & 3.3 (cm/s) \\ \hline
    \small{Compressibility} & ${\cal C}=\frac{\left\langle \left(\stackrel{\rightharpoonup}{\nabla }_{2} \cdot \stackrel{\rightharpoonup}{v}\right)^{2} \right\rangle }{\left\langle \left(\stackrel{\rightharpoonup}{\nabla }_{2} \stackrel{\rightharpoonup}{v}\right)^{2} \right\rangle }$ & 0.49 $\pm$ 0.02\\
    \hline
  \end{tabular}
\end{center}
\end{table}

\section{Information Dimension $D_1$}

The information dimension $D_1$ is measured by dividing the system's field of view into boxes of size $r=s/\eta$, where $s$ is the box size in cm,  making $r$ dimensionless.  One then calculates the probability $P_i(r)$ of box $i$ being populated. The information dimension $D_1$ is defined as:

\begin{equation}
\label{eq_D1}
D_1 =\lim_{r\to 0}\frac{-I(r)}{\log r}
\end{equation}

\begin{figure}
\centerline{\includegraphics[width = 2.8 in]{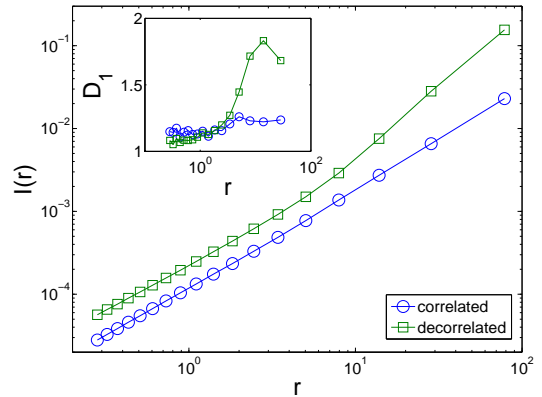}}
\caption{Plot of information function $I(r)$ versus $r=s/\eta$ for the correlated and decorrelated flows at $t/\tau_0=5$. Scale-free behavior is observed for $0.2 < r < 3$. The inset shows the quality of the scale-free behavior. For the correlated flow $D_1=1.16\pm0.02$, and for the decorrelated flow $D_1=1.09\pm0.02$. The most conspicuous difference between the correlated and decorrelated flows is the lack of scaling in the inertial range for the decorrelated flow. This reflects the loss of large-scale structures when the flow is decorrelated.  The prediction for $D_1$ in the Kraichnan ensemble is $D_1 \simeq 1$.}
\label{D1}
\end{figure}

The function $I(r)$, is called the {\em information function}, is defined as

\begin{equation}
\label{eq_I(r)}
I(r) =-\sum_i {P_i(r) \ln (P_i(r))}.
\end{equation}
To calculate $D_1$, one plots $ I(r)$ $vs$ $\log r$. The slope of this line is $D_1$ over the interval of $r$ where this plot is a straight line.  Fig. \ref{D1} shows $I(r)$ vs. $\log r$ of the correlated and decorrelated particle distributions seen in Fig. \ref{particle}. For the correlated flow, $D_1 = 1.16 \pm 0.02$ and for the decorrelated flow, $D_1 = 1.09 \pm0.02$, measured over the interval $0.2<r<3$. We choose to label  this range of $r$ as being in the dissipative range. This is consistent with the fact that the transition from the inertial to dissipative range in many experiments occurs at $r > 1$ \cite{Meneveau96,Benzi93}.

The qualitative impression that both distributions lie along line-like structures is consistent with $D_1$ being close to unity. This finding is consistent with the DNS performed in \cite{BDES2004}, where it was found that $D_1 = 1.15$ for the correlated flow.  At values of  $r>10$ (in the inertial range), the correlated and decorrelated distributions differ significantly. For the correlated flow in Fig. 2 (a), the line-like structures extend over several integral length scales $l_0$. These large-scale structures of the correlated flow give rise to a value of $D_1$ that is larger than its small-scale value ($0.2<r<3$), as more clearly seen in the inset of Fig. 3. This inertial range scaling is consistent with the results of a model for adjustably compressible flow in \cite{Ducasse2008}. For the decorrelated flow, the distribution is more homogeneous over the inertial range, where scale-free behavior is not apparent. This demonstrates (quantitatively) the importance of time correlation for the formation of large-scale coherent structures.

The Kraichnan model of time-decorrelated velocity fluctuations provides a simple relation between the information dimension $D_1$ and the compressibility ${\cal C}$ \cite{BDES2004}:

\begin{equation}
\label{eq_D1(C)}
D_1 =\frac{2}{1+2\cal{C}},
\end{equation}
both of which are measured in this study. The transition to strongly compressibile flow takes place at  ${\cal C} = 1/2$, which is approximately what is seen in this experiment. For the decorrelated flow in this experiment, $D_1 = 1.09 \pm 0.02$, which is measurably greater than unity. The decrease in the dimension $D_1$ for the decorrelated flow in this experiment is consistent with the results in \cite{BDES2004}, but the value obtained is larger than the Kraichnan prediction of $D_1 \simeq 1$.

\section{Lyapunov Exponents}
 The information dimension $D_1$ was obtained in BDES from a measurement of the two Lyapunov exponents, which have been measured in this experiment as well. The Lyapunov exponents of the floaters give a measure of the mean rate at which particle pairs separate on the surface.  In $d$ = 2 dimensions, the largest is $\lambda_1$ (which is positive), and the smallest $\lambda_2 <0$. Since the floaters coagulate, their sum is negative.  These exponents are defined so that each pair is initially separated by a distance $r_0$ that is very small  (less than $\eta$ in the turbulent case).  Since the Lyapunov exponents define an average rate of separation, a pair must be tracked for a long enough time that their rate of separation has become time-independent, but not so long that their final separation has become greater than  $\eta$ \cite{ARTALE1997}.

\begin{figure}
\centerline{\includegraphics[width = 2.8 in]{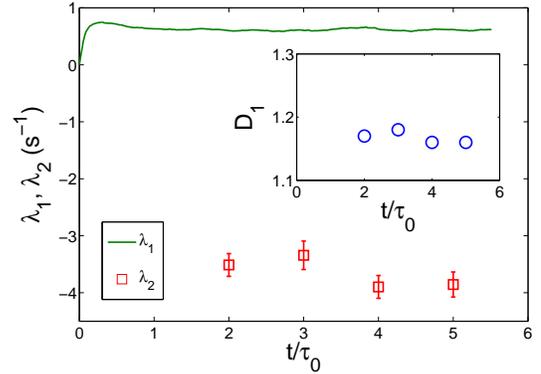}}
\caption{Calculation of the largest Lyapunov exponent $\lambda_1$ for the correlated flow. The information dimension $D_1$ is calculated (inset) at several instants of time once the largest Lyapunov exponent has saturated to a steady-state value of the $\lambda_1=0.6 Hz$. From Eq. \ref{kaplan_yorke}, one can estimate the value of the second Lyapunov exponent $\lambda_2=-3.5\pm0.26 Hz$ from $\lambda_1$ and $D_1$. The sum of the Lyapunov exponents is related to the entropy rate $\dot{S}=\lambda_1+\lambda_2=-2.9 \pm0.3 Hz$. This is comparable to the entropy rate $\dot{S}=-2.4\pm0.02 Hz$ measured in \cite{bandi2006}.}
\label{lambda1_corr}
\end{figure}

\begin{figure}
\centerline{\includegraphics[width = 2.8 in]{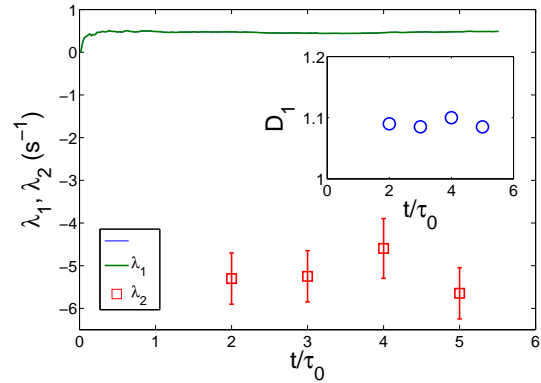}}
\caption{Calculation of the largest Lyapunov exponent $\lambda_1$ for the decorrelated flow. The information dimension $D_1$ is calculated (inset) at several instants of time once the largest Lyapunov exponent has saturated to a steady-state value of the $\lambda_1=0.5 Hz$. From Eq. \ref{kaplan_yorke}, one can estimate the value of the second Lyapunov exponent $\lambda_2=-5.2\pm0.44 Hz$ from $\lambda_1$ and $D_1$. The important difference between the correlated and decorrelated flows is not the change in the Lyapunov exponents, but the decrease in the dimension $D_1$, which is consistent with the results in \cite{BDES2004}}
\label{lambda1_decorr}
\end{figure}

Instead of allowing particle pairs to separate continuously, $\lambda_1$ is calculated by the algorithm given in \cite{Wolf1985}. After each frame, a given particle pair's separation $r(t)$ is reset to $r_0$ and evolved again. This is done in the limit that $t \rightarrow \infty$, where $\lambda_1$ should saturate to it's steady-state limiting value. The particle pairs are evolved using the same Lagrangian evolution scheme discussed above. The results presented here do not depend on $r_0$, which was varied from $0.25 \eta$ to $1 \eta$. An initial separation of $r_0=0.25 \eta$ is used.  Here, we evolve an ensemble of particle pairs ($10^6$) for a maximum of $5 \tau_0$, which corresponds to roughly $100 \tau_{\eta}$ ($\tau_0$ =0.4 s and $\tau_{\eta}$ = 0.04 s). The calculation of $\lambda_1$ is thus:
\begin{equation}
\label{eq_r(r)}
\lambda_1 = \lim_{t\to\infty} \frac{1}{t} \langle \log \frac{r(t)}{r_0} \rangle,
\end{equation}
where the angular brackets denote an ensemble average over the various particle pairs. From Fig. \ref{lambda1_corr} and \ref{lambda1_decorr} it is seen that after a small transient time (of roughly $\tau_{0}$), $\lambda_1$ saturates to a value of 0.6 Hz (correlated) and 0.5 Hz (decorrelated) (see Table \ref{table2}). Also shown in Fig. \ref{lambda1_corr} and \ref{lambda1_decorr} are values of $D_1$ measured at various times during the experiment after $\lambda_1$ has saturated.

One can calculate the value of $\lambda_2$, using $\lambda_1$ and $D_1$, from the Kaplan-Yorke relation (which is exact in two dimensions)\cite{dorfman1999}:
\begin{equation}
\label{kaplan_yorke}
D_1 = 1 + \lambda_1 / |\lambda_2|.
\end{equation}
From this equation, $\lambda_2$ is calculated for both the correlated and decorrelated flows in Fig. \ref{lambda1_corr} and \ref{lambda1_decorr}, respectively. The results from this experiment and from the simulations in BDES, which are summarized in Table \ref{table2}, confirm what the eye sees in Fig. \ref{particle}: $D_1$ is roughly equal to unity (the particles cluster into string-like clusters), and the sum of the Lyapunov exponents is negative.

\begin{table}
\caption{\label{table2}Parameters measured in this experiment, the simulation of BDES, and a prior experiment where $\langle dS/dt \rangle$ was measured \cite{bandi2006}.}
\begin{center}
\begin{tabular}{|p{1.2in}|p{0.4in}|p{0.4in}|p{0.4in}|p{0.4in}|}
 \hline
 & \small{$D_1$} & $\lambda_1$ & $\lambda_2$ & $\langle dS/dt \rangle$ \\ \hline
\small{Experiment}\\ \hline
\small{Correlated} & 1.16 $\pm$ 0.02 & 0.6 $\pm$ 0.05 & -3.75 $\pm$ 0.26 & -3.15 $\pm$ 0.5  \\ \hline
\small{Decorrelated} & 1.09 $\pm$ 0.02 & 0.5 $\pm$ 0.05 & -5.2 $\pm$ 0.44 & -4.7 $\pm$ 0.8  \\ \hline
\small{BDES Simulation}\\ \hline
\small{Correlated} & 1.15  &  &  &  \\ \hline
\small{Decorrelated} & 1.05 &  &  &  \\ \hline
\small{Bandi Experiment} & & & & -2.4 $\pm$ 0.02 \\ \hline
    \hline
  \end{tabular}
\end{center}
\end{table}

There is an additional reason for measuring $\lambda_1$ and $\lambda_2$; their sum is related to a quantity measured in a prior experiment, providing an additional check on the internal consistency of all the experimental observations.
In a prior experiment by Bandi et al. \cite{bandi2006}, the mean rate of change of the entropy of the floaters, $\langle dS/dt \rangle$, was measured in the same tank used in the present experiments, though at a $Re_{\lambda}$ of 100 instead of 160. Here $S$ is the entropy, defined as $S =-\sum_i {P_i \ln (P_i)}$, which is Eq. (\ref{eq_I(r)}) in the limit that the box size $r$ is sufficiently small.

This mean entropy rate $\langle dS/dt \rangle$ is the spatially-averaged divergence of the velocity field at the surface \cite{bandi2006}. It is also equal to the sum of the Lyaunov exponents, $\langle dS/dt \rangle =\lambda_1 + \lambda_2$, if the turbulent flow is statistically stationary \cite{dorfman1999}.  In \cite{bandi2006} $\langle dS/dt \rangle \simeq -2.4$ Hz, while the present experiment gives $\Sigma=-3.15 \pm 0.5$ Hz. In light of the uncertainty in the measurement of the entropy rate in \cite{bandi2006}, and the difference in $Re_{\lambda}$, this agreement between the two types of measurement of $\Sigma$ seems quite satisfactory. In comparing the simulation of BDES with these experiments, it should be born in mind that the laboratory results have appreciable uncertainty, while there is no error (or units) quoted for the simulations.

\section{Summary}

Turbulence is especially difficult to understand because the velocity fluctuations, which lie at its heart, are chaotic in both space and time, with very many degrees of freedom being excited. The phenomenon is especially intractable when viewed in the Eulerian frame, where the small-scale velocity fluctuations are contained in, and moved about, by eddies of larger size.

In an effort to capture the essence of turbulence,  R. Kraichnan, and L. F. Richardson before him, proposed that the problem is more easily understood in the Lagrangian frame, where large-scale sweeping effects are absent.  To effect further simplification, Kraichnan proposed that the temporal velocity fluctuations be treated as in Brownian motion, i.e., that the correlation time of eddies of all sizes, be approximated as being infinitely short (a Gaussian approximation, as in Brownian motion,  is made as well).

This experiment probes a type of compressible flow, namely the chaotic motion of particles that float on an incompressible and strongly turbulent fluid.  The spatio-temporal dynamics of these floaters is studied, and then the observed velocity fluctuations are randomized in time, so as to mimic the simplifying dynamics analyzed by Kraichnan.   As predicted, certain aspects of the particle motion and topology are retained  in the temporally decorrelated field.

The topology of the decorrelated floaters is string-like, just as in the correlated case (compare the two panels in Fig. \ref{particle}.) The information dimension $D_1$ remains close to unity, at least at small spatial scales where this parameter is defined.  On large scales, temporal decorrelation truncates the length of the string-like structures, as seen in Fig. \ref{particle}. The present measurements, and those of BDES \cite{BDES2004}, show that $D_1$ (and presumably other fractal dimensions) is measurably altered by the temporal randomization, but at small scales the particle distribution retains its string-like character.

The effect of temporal randomization on the two Lyaponuv exponents is more delicate.  For both the correlated and decorrelated flows,  the positive exponent $\lambda_1$ remains  positive (patches of particles are stretched into strings in both cases), but decorrelation compacts the structures into thinner strings, i.e., $\lambda_2$ is decreases from roughly -3 to -5 Hz.

The steady-state entropy rate of change $dS/dt$, which for a flow admitting SRB statistics is the sum of $\lambda_1$ and $\lambda_2$ \cite{falkovicha,falkovichb}, is lowered by decorrelation; it becomes even more negative.  The measurements reported here, and the simulations of Boffetta et al., suggest that our understanding of turbulence can be enhanced by comparing and contrasting correlated and temporally decorrelated flows.
\section{Acknowledgements}
We acknowledge very helpful discussions with M.M. Bandi, A. Pumir, G.Boffetta, and J. Schumacher.
Funding was provided by the US National Science Foundation grant
$\#$ DMR-0604477.
\bibliography{all}

\end{document}